\documentclass[12pt]{iopart}

\usepackage[dvips]{graphicx}

\begin{document}

\title[A tunable rf SQUID manipulated as flux and phase qubit]
{A tunable rf SQUID manipulated as flux and phase qubit}

\author{S Poletto$^1$, F Chiarello$^2$, M G Castellano$^2$, J Lisenfeld$^1$,
A Lukashenko$^1$, P Carelli$^3$ and
A V Ustinov$^1$\footnote{Author to whom any correspondence should be
addressed.}}

\address{$^1$ Physikalisches Institut, Universit{\"a}t
Karlsruhe (TH), D-76131 Karlsruhe, Germany}
\address{$^2$ Istituto di Fotonica e Nanotecnologie, CNR, 00156 Roma, Italy}
\address{$^3$ Dip. Ingegneria Elettrica, Universit\`a dell'Aquila,
67040 Monteluco di Roio, Italy}

\ead{ustinov@physik.uni-karlsruhe.de}

\begin{abstract}
We report on two different manipulation procedures of a tunable rf
SQUID. First, we operate this system as a flux qubit, where the
coherent evolution between the two flux states is induced by a rapid
change of the energy potential, turning it from a double well into a
single well. The measured coherent Larmor-like oscillation of the
retrapping probability in one of the wells has a frequency ranging
from 6 to 20 GHz, with a theoretically expected upper limit of 40
GHz. Furthermore, here we also report a manipulation of the same
device as a phase qubit. In the phase regime, the manipulation of the
energy states is realized by applying a resonant microwave drive. In
spite of the conceptual difference between these two manipulation
procedures, the measured decay times of Larmor oscillation and
microwave-driven Rabi oscillation are rather similar. Due to the
higher frequency of the Larmor oscillations, the microwave-free
qubit manipulation allows for much faster coherent operations.
\end{abstract}


\maketitle

\section{Introduction}

Superconducting qubits are promising systems for the realization of
quantum computation. Their coherent evolution, entanglement, storage
and transfer of quantum information as well as quantum non-demolition readout
are just few examples of what has been already achieved
\cite{Mooij-03, Berkley-02, Simmonds-07, Mooij-06}. One of the major
challenges superconducting qubits are facing now is to increase
their coherence time in order to reach an useful number of quantum
operations \cite{knill-05}. However, there is an alternative
possibility of achieving the same goal: one can try to increase the
\textit{qubit operation speed} to make qubit gates shorter, thus
leading to less restrictive requirements for the coherence time.
Nowadays, qubit operation frequencies exceeding 500 MHz are
difficult to reach with phase qubits manipulated by microwave
signals, since the limited anharmonicity of the qubit energy potential
gives rise to a leakage into higher excited states.

In this paper, we report experiments with a tunable rf SQUID and
demonstrate two different manipulation procedures. The system under
test, a double SQUID, can be operated in two different regimes: (i)
as a flux qubit manipulated via fast (dc) pulses of magnetic flux,
and (ii) as a phase qubit in which the quantum state evolution is
controlled by microwave pulses of chosen amplitude and phase. The
former manipulation approach without using microwaves allows to reach
very high oscillation frequencies, while operating the
same circuit as a phase qubit offers the possibility to verify the
obtained results using a well defined and studied
manipulation technique. We will see in the following that the
coherence times measured using the two different manipulation
schemes are rather similar, suggesting that decoherence acts in
a similar way in both cases. This conclusion draws attention to
common sources of decoherence (presumably, dielectric loss due to
two-levels fluctuators in the material used for chip fabrication
\cite{Martinis-05, McDermott-09}), and emphasizes the relevance of
increasing the number of qubit rotations within the coherence time
by increasing the oscillation frequency of the system.

The paper is organized as follows. A detailed overview of the system
under test, composed of the double SQUID and the readout dc SQUID,
is given in \sref{sec:setup}. \Sref{sec:mw-free} describes the
manipulation of the double SQUID by deforming its energy potential
using fast pulses of magnetic flux. The measured coherent
oscillations are presented here together with their theoretical
interpretation. \Sref{sec:mw-manipulation} reports the manipulation
of the same device as a phase qubit. Rabi oscillations measured
via microwave pulses of variable length are reported. The dependence
of the Rabi oscillation frequency on microwave power is analyzed
by taking into account the population of higher excited states.

\section{System and experimental setup}\label{sec:setup}

\begin{figure}
\begin{center}
\includegraphics{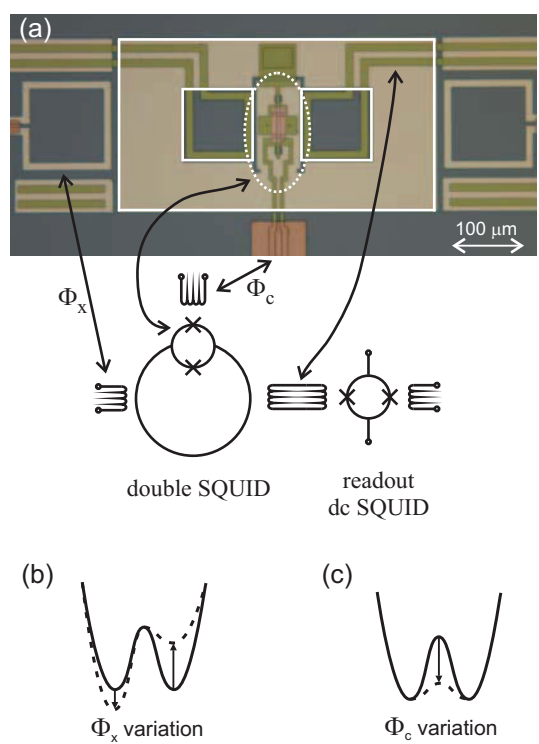}
\end{center}
\caption{(a) Photo and schematic representation of the studied double SQUID.
        The entire device occupies a space of $430\times 230$ $\mu \mathrm{m}^2$.
        (b)-(c) Manipulation of the energy potential profile via
        the two bias fluxes $\Phi_x$ and $\Phi_c$. Variation of the symmetry
        is achieved by changes of the flux $\Phi_x$, while the barrier
        between the two local minima is tuned by the flux $\Phi_c$.}
\label{fig:1}
\end{figure}

The circuit that we studied consists of a superconducting Nb loop of
inductance $L=85$ pH interrupted by a small dc SQUID of inductance
$l=6$ pH. The name for this kind of device varies between different
authors from 'double SQUID' \cite{Chiarello-05} to 'combined rf-dc
SQUID' \cite{Han-89} or 'modified rf SQUID' \cite{Bennett-08}. The
two Josephson junctions embedded into the dc SQUID are nominally
identical except for unavoidable asymmetries originating in the
fabrication process. Each junction has a critical current of 8
$\mu$A and a self capacitance of 0.4 pF. The system is manipulated
via two magnetic bias fluxes $\Phi_x$ and $\Phi_c$ applied to the
large and small loops, respectively. The detailed characterization
of the device taking into account non-identical Josephson junctions and
non-negligible inductance $l$ has been reported elsewhere
\cite{Castellano-07}. Both loops are designed fully gradiometrically
with the intent to decrease both the noise induced by external
uniform magnetic fields as well as cross talk between the two bias fluxes.
A photograph of the double SQUID together with its schematic
representation is presented in \fref{fig:1}(a). The
double SQUID is defined by the pale-white area delimited by
white solid lines at the center of the picture.
The two large holes of $100\times 100$ $\mu \mathrm{m}^2$
define the gradiometric main loop. In the pale-white area
highlighted by the white dashed ellipse one can identify
two much smaller holes
($10\times 10$ $\mu \mathrm{m}^2$), defining the gradiometric inner
dc SQUID. The two Josephson junctions, of dimensions $3 \times 3
\,\mu m^2$, are visible at the center. The two coils on the left and
right side are used to control the bias flux $\Phi_x$; the
mutual inductance between them and the double SQUID is 2.6 pH. The
coil inducing the bias flux $\Phi_c$, visible on the lower-central
part of the picture, is wrapped around one of the two small loops,
to which it has a mutual inductance of 6.3 pH. The placement of coils and
holes reduces the cross talk between $\Phi_x$ and $\Phi_c$ lines to
less than 1\%, as verified experimentally. Each of the two coils
inside the large loops forms a part of a superconducting transformer
connecting the system to an unshunted readout dc SQUID (one for each
side). The circuit was made by Hypres \cite{Hypres} using standard
Nb/AlO$_x$/Nb technology, with a critical current density of 100
A/cm$^2$ and SiO$_2$ as dielectric material for junction isolation.

The inductance of the small loop $l$ is chosen to be much smaller
than that of the main loop $L$, so that the two-dimensional energy
potential defining the system can be approximated by a
one-dimensional function of the parameter $\delta$, corresponding to
the phase difference across the inner dc SQUID. Moreover, the
inductance of the main loop $L$ and the critical current of each
junction $I_0$, are chosen such that the parameter $\beta_L = 2\pi
L\cdot 2I_0/\Phi_0$ ranges between 1 and $5\pi/2$, resulting in a
double well potential. The two bias fluxes $\Phi_x$ and $\Phi_c$ are
used to manipulate the energy potential profile. Changes on $\Phi_x$
modify the symmetry (\fref{fig:1}(b)) of the potential, while
changes on $\Phi_c$ tune the height of the barrier between the two
local minima (\fref{fig:1}(c)).

\begin{figure}
\begin{center}
\includegraphics{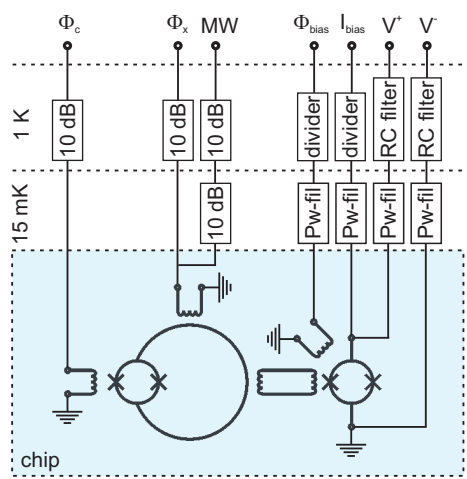}
\end{center}
\caption{Representation of the experimental setup. The pale-blue area
        indicates the device on chip, while the symbols \textit{Pw-fil},
        \textit{divider}, \textit{RC filter} and \textit{10 dB} indicate
        powder filters, current dividers, RC filters and attenuation
        values on coax lines.}
\label{fig:experimental_setup}
\end{figure}

Measurements were performed at the base temperature of a dilution
fridge stabilized at 15 mK. The experimental setup is reported in
\fref{fig:experimental_setup}. The two currents generating the fluxes
$\Phi_x$ and $\Phi_c$ were supplied via coaxial cables with 10 dB
attenuators anchored at the 1K-pot stage of the refrigerator.
The microwave signal, applied to the double SQUID via the coil
$\Phi_x$, was supplied through an extra coaxial cable
with 10 dB attenuation at the 1K-pot stage and 10 dB
at base temperature.
For biasing the readout dc SQUID we used superconducting wires,
metal powder filters \cite{Lukashenko-08} at base temperature and
current dividers at the 1K-pot stage, while the lines to measure the
voltage response were equipped with low pass filters instead of current
dividers at the 1K-pot stage.

\section{Microwave-free manipulation}\label{sec:mw-free}

The qubit manipulating scheme reported in this section is based on
changing the double well potential to a single well shape and back,
caused by fast dc pulses on the bias flux $\Phi_c$. Hereby, the
computational states are mapped to the flux states of the double
well potential (i.e., left and right wells), while the coherent
evolution occurs between the two lowest energy eigenstates of the
single well potential.

\begin{figure}
\begin{center}
\includegraphics{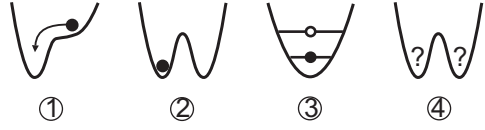}
\end{center}
\caption{Variation of the potential shape in the \textit{microwave-free}
         manipulation of the double SQUID.}
\label{fig:2}
\end{figure}

The manipulation procedure is described in detail in
Ref.~\cite{Poletto-09}. It consists in the four main steps depicted
in \fref{fig:2}(c) and outlined as follows.
\begin{itemize}

\item [\textcircled{1}] The state of the qubit is initialized. A pulse on the flux $\Phi_x$ tilts the potential in order to remove one of the two minima. This strongly asymmetric potential is maintained for a time required for the complete relaxation in the energy minimum.

\item  [\textcircled{2}]  After the initial qubit state is prepared, the potential is turned into the symmetric double well. The qubit state remains frozen due to the large barrier height between the wells, which prevents any tunneling between them. In this case, only pure dephasing is possible as only one of the two states is populated. Which of the two wells is populated depends on the sign of the flux pulse in the previous step.

\item  [\textcircled{3}] Here we perform the quantum manipulation: By applying a short flux pulse $\Phi_c$ the barrier is completely removed and the potential is turned from the symmetric double well into single well. This deep single well  can be approximated by a parabola of a harmonic oscillator, whose characteristic frequency depends on the pulse height. When the pulse is over, the potential returns to the initial deep double-well configuration.

\item  [\textcircled{4}]  The final circuit state is read out by applying a current ramp through the dc SQUID and recording its switching current  to the non-zero voltage state.

\end{itemize}

\begin{figure}
\begin{center}
\includegraphics{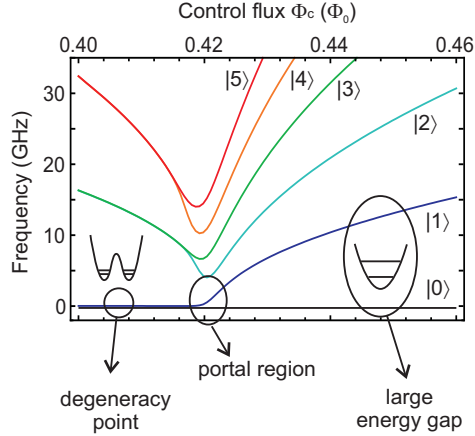}
\end{center}
\caption{Energy values of the first six eigenvalues for the double-SQUID.
        The ground state $|0\rangle$ is defined to be the zero reference
        level.}
\label{fig:energy_levels}
\end{figure}

The main point of above procedure is the non-adiabatic transition
between the lowest two energy levels, which occurs during the potential
transformation. In \fref{fig:energy_levels} the energy values, expressed as a
frequency, of the first six eigenvalues for the double-SQUID are reported
versus the control flux $\Phi_c$. On that graph the energy of the ground
state $|0\rangle$ is defined to be the zero reference level.
Symmetric double well situation and single well situation are depicted,
respectively, on the left and right side of the graph.

For a perfectly symmetric double well, the lowest
two energy levels are degenerate, with an energy splitting $\Delta$
between them depending exponentially on the inverse of the barrier height.
For the more realistic case of non perfect symmetry, we express the
asymmetry $\epsilon$ as the energy difference between the two minima of the potential.
For large potential barrier, the level splitting
$\hbar\Omega = \sqrt{\epsilon^2+\Delta^2}$ is dominated by $\epsilon$ and
remains essentially constant during the initial drop of the barrier.

When the barrier is almost completely removed (for $\epsilon \sim \Delta$,
which is indicated in \fref{fig:energy_levels} as "portal" region \cite{Koch-06}),
the level splitting starts to increase and saturates at the single-well
oscillation frequency. This process corresponds to "half" a Landau-Zener process,
starting from the degeneracy point and arriving at the large energy gap
condition (see \fref{fig:energy_levels}).
In this process, the system is prepared and remains initially in a 50\%
superposition of the two lowest energy eigenstates. The initial phase of
this superposition depends on the chosen initial state of the deep double-well
potential. For example, an initial left $|L\rangle$ (right $|R\rangle$)
flux state is transformed in a 50\% superposition with positive (negative) sign.
Although unwanted non-adiabatic transitions to upper levels (from the third onward)
are possible, they can be suppressed owing to the gap existing between
the first two levels and upper levels in the portal region.
This requires an appropriate choice of the pulse rise/fall-times so that
the transition is performed non-adiabatically for the computational
state but adiabatically for upper levels.

After crossing the portal region, the system is maintained for the
pulse duration in the deep single-well potential. Here, the lowest
two levels are equally populated and the relative phase between them
evolves with a rate given by the level splitting. This level splitting
is tunable by the flux pulse amplitude, and can be described by the formula
\begin{equation}\label{eq:frequency_vs_phic}
\Delta E=\hbar\omega_0(\Phi_c)\approx\frac{\hbar}{\sqrt{2LC}}\sqrt{1-\beta(\Phi_c)}
\end{equation}
where $\beta(\Phi_c)=\beta_L\cos(\pi\Phi_c/\Phi_0)$ is the modification
of the parameter $\beta_L$ via the flux $\Phi_c$.
During this time the system is weakly responsive to fluctuations on the
flux $\Phi_c$ and so it is naturally protected against noise\footnote{as an
example, a fluctuation of $1\mu\Phi_0$ on the flux $\Phi_c$ is responsible
of a change in the oscillation frequency of only 125 kHz, corresponding
to a percentage variation of $6.6\times10^{-4}$ at 19 GHz.}.
The final phase between the states is determined by the duration of the
flux pulse, i.e. by the time elapsed in the single well condition.
At the end of the pulse the flux returns to the initial condition
and the system goes back to the two well state. The portal region is crossed again,
and the inverse of the previous process occurs: the relative phase
between the two energy states is transformed to the amplitude of the
left/right states. After the pulse, the system's flux state becomes
once again frozen, and it is read out by an unshunted dc SQUID magnetometer
which is weakly inductively coupled to the qubit loop.

\begin{figure}
\begin{center}
\includegraphics{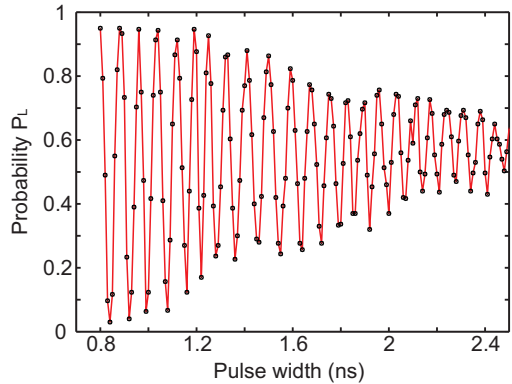}
\end{center}
\caption{Probability of measuring the state $|L\rangle$ as a function of the pulse duration. The coherent oscillation shown here has a frequency of 14 GHz and a coherence time of approximately 1.2 ns.}
\label{fig:3}
\end{figure}

\begin{figure}
\begin{center}
\includegraphics{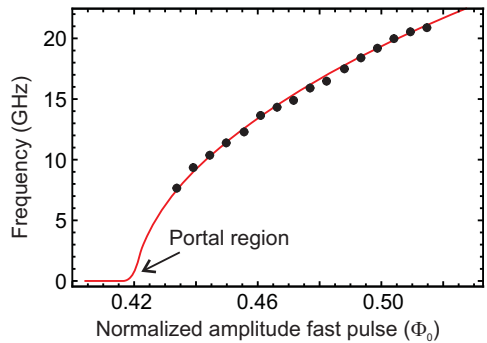}
\end{center}
\caption{Measured oscillation frequencies versus amplitude of the short
         flux pulse (full dots). The solid curve is a numerical simulation using
         the measured parameters of the circuit.}
\label{fig:4}
\end{figure}

The described procedure was repeated many times (from 100 to 10000) in order to determine the probability of one of the projected states for a chosen combination of parameters, such as pulse height and duration. This is then repeated for different pulse durations $\Delta t$ in order to record the coherent oscillations of the qubit state (\fref{fig:3}). From curves collected for different pulse amplitudes we extracted the oscillation frequency and the decay time. In \fref{fig:4} we show the measured oscillation frequency (full dots) versus the pulse amplitude. These data are in very good agreement with the theoretical expectation given by \eref{eq:frequency_vs_phic} for the single-well oscillator frequency (solid line). The measured decay time is on the order of a nanosecond, independent of the frequency.

The described manipulation scheme only allows for $R_x$-rotations
around the qubit's Bloch sphere. The full qubit control requires
also phase control achieved by $R_z$-rotations. These, in principle,
can be realized by slightly unbalancing the potential through a
dc pulse on the bias flux $\Phi_x$.
Quantum computation protocols require also the interaction between
different qubits  which can be envisaged by using an inductive
switchable coupling that allows for $\sigma_z\sigma_z$ coupling
between them \cite{Chiarello-00}.

It is worth noting that the first manipulation of a superconducting
qubit with fast dc pulses was achieved by Nakamura and collaborators
on a Cooper-pair box in 1999 \cite{Nakamura-99}.
In that article, coherent oscillations of frequencies up to 19 GHz
are reported. In spite of the similar manipulation technique the
coherent oscillations reported by us are due to a different physical process.
In the charge qubit, the manipulation was done at the degeneracy point
where ground and first excite state, defined by $n$ and $n+1$ cooper
pairs on the island, are no eigenstates of the energy.
The population of the initially prepared ground state is thus
oscillating with the first excited state due to the off-diagonal terms
in the Hamiltonian. The coherent oscillations reported in \cite{Nakamura-99}
are a direct measurement of the population evolution.
In our system the free evolution is performed in a single well
potential where there are no degeneracy points. The phase evolution of
each populated energy level is the only physical process taking place.
The phase difference between ground and first excite state is not
detectable in the energy base, but it can be easily measured by projecting
the state of the system in the flux base defined by the double well situation.
The reported coherent oscillations are thus a direct consequence of the
not directly measurable phase evolution in the energy base.

\section{Microwave-induced manipulation}\label{sec:mw-manipulation}

In addition the above microwave-free manipulation, we have also operated the system as a phase qubit using microwave driving. The operating procedure is the same as that reported in Ref.~\cite{Lisenfeld-07}. The energy potential profile is strongly tilted via the flux bias $\Phi_x$ making one of the two local minima shallow enough to contain only a small number of energy levels. The two computational states of the qubit are defined to be ground $|0\rangle$ and first excited states $|1\rangle$ in the shallow well. The first excited state is populated by resonant absorption of photons from the microwave field, while the complete manipulation on the Bloch sphere can be performed via microwave pulses of defined duration and phase and dc flux pulses. The main difference between our device and the conventional rf SQUID phase qubits \cite{Steffen-06} is the possibility to tune \textit{in situ} the Josephson energy of the device via the flux coil $\Phi_c$.
This additional tuning parameter allows to modify the anharmonicity
of the energy potential in a slightly different way to what is done by
the flux $\Phi_x$. It is thus possible to minimize the leakage to higher
excited states in a strongly deformed potential (3-4 energy levels in
the shallow well) without changing the number of levels inside
the well nor the escape probability of each
state.
The readout procedure is performed in two steps:

\begin{enumerate}

\item An adiabatic, but fast, dc pulse is sent to the bias coil $\Phi_x$,
    the potential is thus deformed and consequently the barrier separating
    the two wells is reduced. The amplitude of the pulse is calibrated
    such that the transition from the shallow to the deep well is triggered
    for the ground state $|0\rangle$ with a probability of approximately
    10\%. That ensure a complete escape of the system from the first excited
    state $|1\rangle$, leading to a theoretical visibility close to 90\%.

\item The flux of the double SQUID is measured with the dc SQUID coupled
    inductively to it, using the procedure already described in \sref{sec:mw-free}.

\end{enumerate}

\begin{figure}
\begin{center}
\includegraphics{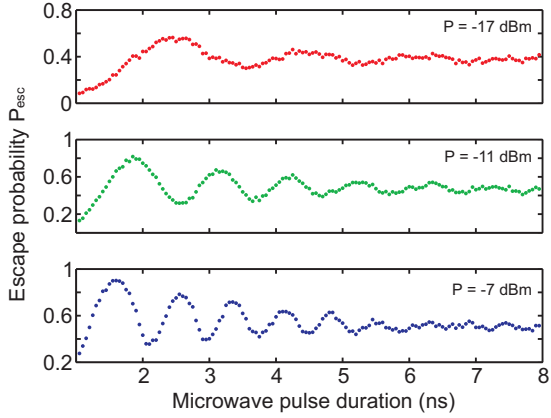}
\end{center}
\caption{Rabi oscillation of the double SQUID manipulated as a phase qubit by applying microwave pulses at 19 GHz. The oscillation frequency changes from 540 MHz to 1.2 GHz by increasing the power of the microwave signal by 10 dB.}
\label{fig:5}
\end{figure}

The sequence of preparation, manipulation and readout is repeated
between 100 to 10000 times, depending on the desired statistical error.
An example of three Rabi oscillations obtained for different powers
of the microwave driving field at 19 GHz is reported in \fref{fig:5}.
The increase of the oscillation frequency with microwave power is
clearly seen. The topmost oscillation has a frequency of $f_{\rm Rabi}=540$ MHz.
Oscillations at lower frequencies are difficult to measure since the
decay time is very short, between 1.5 and 2.0 ns. On the lower graph,
the oscillation frequency is approximately equal to $f_{\rm Rabi}=1.2$ GHz.
Further increasing the power leads, as we will see, to an unwanted
population of higher excited states.

\begin{figure}
\begin{center}
\includegraphics{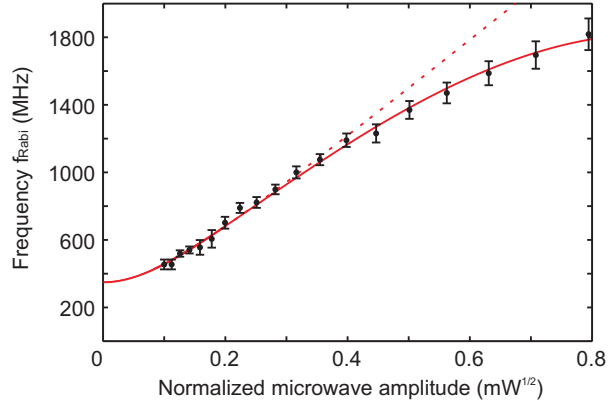}
\end{center}
\caption{Measured Rabi oscillation frequency versus the normalized amplitude of the microwave signal (solid dots). The dashed line is a linear
fit taking into account slightly off-resonance microwave field, while the fit represented by the solid line considers a population of higher excited states.}
\label{fig:6}
\end{figure}

The correlation of the Rabi oscillation frequency versus amplitude
of the microwave driving field is reported in \fref{fig:6}.
The deviation from the linearity visible at higher powers is a clear
indication of populating higher excited states \cite{Claudon-08}.
We note also that the qubit has have been driven slightly off-resonance since the frequency of Rabi oscillations does not reach zero for zero power.

The experimental data in \fref{fig:6} were fitted with two curves:

\begin{itemize}

\item The dashed line represents a linear power-frequency relation with
      the addition of a term taking into account the microwave pumping
      out of resonance. The equation has the form
      $$
      f_{\rm Rabi}=\sqrt{(f_{01}-f_{\rm MW})^2 + {\left(k A\right)}^2}
      $$
      where $A$ is the normalized amplitude of the microwave driving signal
      in mW$^{1/2}$ and $k$ is  a constant.  The microwave frequency
      $f_{\rm MW}=19$ GHz is off set from the resonance frequency $f_{01}$
      corresponding to the transition between the lowest two levels.
      The points used for the fit refer to a normalized microwave amplitude
      ranging between 0.1 $mW^{1/2}$ and 0.4 $mW^{1/2}$.
      From the fit we obtained a value of $k= (2919\pm 105)$ MHz/mW$^{1/2}$
      and $f_{01}-f_{\rm MW}=(349 \pm 45)$ MHz.

\item The solid fitting curve takes into account both a non-zero
      population of higher excited states due to large driving field
      and the manipulation red-detuned from the resonance.
      The dependence of the Rabi frequency on the power of the microwave
      signal is described e.g. in Ref.~\cite{Amin-04} (equation 22).
      The function that we used for the solid curve fit has the form
      $$
      f_{\rm Rabi}=\sqrt{(f_{01}-f_{\rm MW})^2 + {\left(k A\left(1-\beta A^2 \right)\right)}^2}.
      $$
      As before, $(f_{01}-f_{\rm MW})$ is the frequency out of resonance
      and $k$ the linear power-frequency correlation, while the new
      coefficient $\beta$ depends on the details of the system. The result
      of the fit, this time made through all measured points, gives the
      values of $k= (2977\pm 104)$ MHz/mW$^{1/2}$, $f_{01}-f_{\rm MW}=(350\pm 42)$ MHz
      and $\beta = 0.411\pm 0.060$ mW$^{-1}$. Within the noted errors,
      both fits yield the same values for the common parameters.

\end{itemize}

\begin{figure}
\begin{center}
\includegraphics{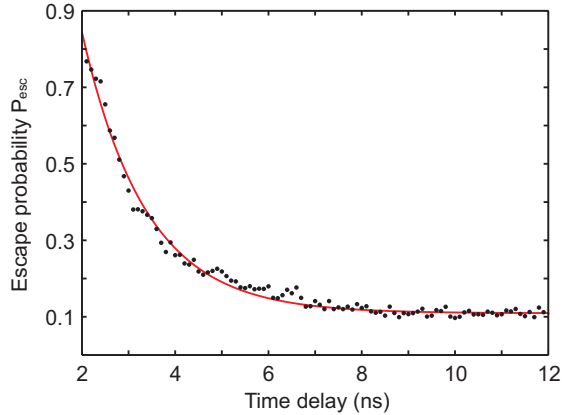}
\end{center}
\caption{Measurement of the relaxation time $T_1$ for the double SQUID
         operated as a phase qubit.}
\label{fig:7}
\end{figure}

We measured the energy relaxation time $T_1$ of the double SQUID when it was operated as a phase qubit. The occupation probability of the first excited state was measured after a variable time between a resonant microwave $\pi$-pulse and the readout pulse. The measured probability decay was fitted exponentially yielding $T_1 = 1.37$ ns (\fref{fig:7}).

\section{Conclusion}\label{sec:conclusion}

We presented measurements on a double SQUID manipulated both as a double-well flux qubit and as a phase qubit. The device manipulation as a flux qubit was performed by modification of its energy potential profile via fast dc pulses, while the manipulation as a phase qubit required the use of microwave signals to induce Rabi oscillations between the ground and first excited state of a shallow well in a strongly deformed potential. The measured coherence time of the Larmor oscillations obtained in the flux regime is about 1 ns, of the same order as the relaxation time $T_1$ in the phase regime. Since best available phase qubits already display relaxation times of several 100 ns, obtained by using appropriate materials in the fabrication processes \cite{Martinis-05}, we suppose that also the coherence time of the Larmor oscillations obtained without microwaves could strongly benefit from the same treatment. Such a possible improvement of the coherence by two orders of magnitude, together with the much higher oscillation frequency in the microwave-free Larmor mode, should in principle allow to reach the ultimate goal of $10^{4}$ single-qubit gate operations within the coherence time which is needed for the implementation of quantum algorithms.

\ack

This work was supported by the the Deutsche Forschungsgemeinschaft (DFG) and the EU project EuroSQIP.


\section*{References}
\begin{thebibliography}{99}

\bibitem{Mooij-03} Chiorescu I, Nakamura Y, Harmans C J P M and Mooij J E 2003
 Coherent quantum dynamics of a superconducting flux qubit
 \textit{Science} \textbf{299} 1869

\bibitem{Berkley-02} Berkley A J, Xu H, Ramos R C, Gubrud M A, Strauch F W,
Johnson P R,
 Anderson J R, Dragt A J, Lobb C J and Wellstood F C 2003
 Entangled macroscopic quantum states in two supereconducting qubits
 \textit{Science} \textbf{300} 1548

\bibitem{Simmonds-07} Sillanp\"a\"a M A, Park J I and Simmonds R W 2007
 Coherent quantum state storage and transfer between two phase qubits via a
resonant cavity
 \textit{Nature} \textbf{449} 438

\bibitem{Mooij-06} Lupa\c{s}cu A, Driessen E F C, Roschier L, Harmans C J P M
and Mooij J E 2006
 High-contrast dispersive readout of a superconducting flux qubit using a
monlinear resonator
 \PRL \textbf{96} 127003

\bibitem{knill-05} Knill E 2005
 Quantum computing with realistically noisy devices
 \textit{Nature} \textbf{434} 39

\bibitem{Martinis-05} Martinis J M \textit{et al} 2005
 Decoherence in Josephson qubits from dielectric loss
 \PRL \textbf{95} 210503

\bibitem{McDermott-09} McDermott R 2009
 Material origins of decoherence in superconducting qubits
 \textit{IEEE Trans. Appl. Supercond.} \textbf{19} 2

\bibitem{Chiarello-05} Chiarello F, Carelli P, Castellano M G, Cosmelli C,
Gangemi L,
 Leoni R, Poletto S, Simeone D and Torrioli G 2005
 Superconducting tunable flux qubit with direct readout scheme
 \textit{Supercond. Sci. Technol.} \textbf{18} 1370

\bibitem{Han-89} Han S, Lapointe J and Lukens J E 1989
 Thermal activation in a two-dimensional potential
 \PRL \textbf{63} 1712

\bibitem{Bennett-08} Bennett D A, Longobardi L, Patel V, Chen W, Averin D V and
Lukens J E 2009
 Decoherence in rf SQUID qubits
 \emph{Quant. Inform. Process.} \textbf{8} 217

\bibitem{Castellano-07} Castellano M G \textit{et al} 2007
 Catastrophe observation in a Josephson-junction system
 \PRL \textbf{98} 177002

\bibitem{Hypres} Hypres Inc., Elmsford, N.Y., USA.

\bibitem{Lukashenko-08} Lukashenko A and Ustinov A V 2008
 Improved powder filters for qubit measurements
 \RSI \textbf{79} 014701

\bibitem{Poletto-09} Poletto S, Chiarello F, Castellano M G, Lisenfeld J,
Lukashenko A,
 Cosmelli C, Torrioli G, Carelli P and Ustinov A V 2009
 Coherent oscillations in a superconducting tunable flux qubit manipulated without microwaves
 \NJP \textbf{11} 013009

\bibitem{Koch-06} Koch R H, Keefe G A, Milliken F P, Rozen G R, Tsuei C C, Kirtley J R and DiVincenzo D P 2006 Experimental Demonstration of an Oscillator Stabilized Josephson Flux Qubit \PRL \textbf{96} 127001

\bibitem{Chiarello-00} Chiarello F 2000
 Quantum computing with superconducting quantum interference devices: a possible strategy
 \PL A \textbf{277} 189

\bibitem{Nakamura-99} Nakamura Y, Pashkin Yu A and Tsai J S 1999
    Coherent control of macroscopic quantum states in a single-Cooper-pair box
    \textit{Nature} \textbf{398} 786

\bibitem{Lisenfeld-07} Lisenfeld J, Lukashenko A, Ansmann M, Martinis J M and Ustinov A V 2007
 Temperature dependence of coherent oscillations in Josephson phase qubits  \PRL \textbf{99} 170504

\bibitem{Steffen-06} Steffen M, Ansmann M, McDermott R, Katz N, Bialczak R C, Lucero E,  Neeley M, Weig E M, Cleland A N and Martinis J M 2006 State tomography of capacitively shunted phase qubits with high fidelity
 \PRL \textbf{97} 050502

\bibitem{Claudon-08}  Claudon J, Zazunov A, Hekking F W J and Buisson O 2008 Rabi-like oscillations of an anharmonic oscillator: classical versus quantum interpretation \emph{Phys. Rev. B} \textbf{78} 184503

\bibitem{Amin-04} Amin M H S 2006
 Rabi oscillations in systems with small anharmonicity
 \textit{Low. Temp. Phys.} \textbf{32} 198


\end{thebibliography}
\end{document}